\documentclass[journal,onecolumn]{IEEEtran}

\usepackage[pdftex]{graphicx}
\usepackage{amsmath}
\usepackage{graphicx}
\usepackage{enumerate}
\usepackage{xcolor}
\usepackage{subfigure}
\usepackage{bm}
\usepackage{tabularx}
\usepackage{tabu}
\usepackage{multicol}
\usepackage{multirow}
\usepackage{colortbl,booktabs}
\usepackage{dcolumn}
\usepackage{algorithm}
\usepackage{algpseudocode}
\usepackage{url}

\graphicspath{{Figures/}}

\newtheorem{theorem}{Theorem}
\newtheorem{remark}{Remark}
\newtheorem{definition}{Definition}
\newtheorem{proof}{Proof}

\newcommand{\bmh}[1]{\hat{\bm{#1}}}

\begin{document}

\title{Why Are the ARIMA and SARIMA not Sufficient}
\author{Shixiong~Wang,
        Chongshou~Li,
        and Andrew~Lim
\thanks{S. Wang and A. Lim are with the Department of Industrial Systems Engineering and Management, National University of Singapore. E-mail: s.wang@u.nus.edu; isealim@nus.edu.sg.}
\thanks{Chongshou Li was with the Department of Industrial Systems Engineering
and Management, National University of Singapore, Singapore 117576. He is
now with the School of Information Science and Technology, Institute of
Artiﬁcial Intelligence, Southwest Jiaotong University, Chengdu 611756, China
(e-mail: cslichina2020@outlook.com).
}
\thanks{This work is supported by National Research Foundation of Singapore under grant NRF-RSS2016-004.}
}

\maketitle

\begin{abstract}
The autoregressive moving average (ARMA) model takes the significant position in time series analysis for a wide-sense stationary time series. The difference operator and seasonal difference operator, which are bases of ARIMA and SARIMA (Seasonal ARIMA), respectively, were introduced to remove the trend and seasonal component so that the original non-stationary time series could be transformed into a wide-sense stationary one, which could then be handled by Box-Jenkins methodology. However, such difference operators are more practical experiences than exact theories by now. In this paper, we investigate the power of the (seasonal) difference operator from the perspective of spectral analysis, linear system theory and digital filtering, and point out the characteristics and limitations of (seasonal) difference operator. Besides, the general method that transforms a non-stationary (the non-stationarity in the mean sense) stochastic process to be wide-sense stationary will be presented.
\end{abstract}


\IEEEpeerreviewmaketitle

\section{Introduction}
One of the intriguing topics of time series analysis is to physically analyse the internal mechanism/dynamics of a system generating the focused time series, and subsequently build a proper mathematical model to describe the dynamics of this system so that we can predict the future with satisfying accuracy. Generally such a time series is a stochastic process rather than a deterministic one which makes the problem more complex.

When it comes to the stochastic process modeling, the reputed Box-Jenkins methodology \cite{box2015time}, also known as ARMA and ARIMA model, stands out. The philosophy of the ARMA model is from the Wold's Decomposition theorem \cite{papoulis2002probability}. The theorem supports that the ARMA model is mathematically sufficient to describe a \textbf{regular} wide-sense stationary (WSS) stochastic process. After the modeling, the least square method, maximum likelihood method and spectral estimation method et al. could be utilised to estimate the parameters of the model based on the collected time series samples.  As a result, we could make use of the past information (i.e., collected samples) to reconstruct the underlying dynamics of the focused stochastic process, and further make satisfying prediction. As complements to ARMA, the ARIMA (resp.  SARIMA) aims to transform the focused non-stationary (in the mean sense) stochastic process to be stationary by difference (resp. seasonal difference) operator with proper order so that the resulted time series could be fed into an ARMA model. Therefore, in Box-Jenkins methodology, the predictable component of a WSS process in one realization is treated as the trend of the regular component of the same WSS process which can be eliminated by the difference or seasonal difference operator. For notation breifness, we collectively refer to ARIMA and SARIMA as S-ARIMA.

However, such difference operators are more empirical experiences than exact theories. Therefore, we do not know why they work well somewhere and ineffectively elsewhere. To this end, in this paper, we aim to investigate the power of the (seasonal) difference operator from the perspective of spectral analysis, linear system theory and digital filtering, and point out the characteristics and limitations that (seasonal) difference operator and S-ARIMA hold. Besides, the general operator that works for transforming a non-stationary (in the mean sense) stochastic process to be wide-sense stationary will be presented. At last, we will show the overall methodology for predicting a non-stationary stochastic process, which is the generalization of S-ARIMA and termed as ARMA-SIN. Our discusses will be based on Linear System Theory, Spectral Analysis, and Digital Filtering. For preliminaries on those topics, refer to  \cite{diniz2010digital,yang2009signals,chaparro2018signals}.

Section \ref{sec:notation} defines some notations.
In Section \ref{subsec:Wold's-Decomposition}, we review some preliminaries about stochastic process and explain the rationale behind the Box-Jenkins methodology (i.e., the S-ARIMA method).
In Section \ref{sec:warming-up}, some examples are given to show the insufficiencies of the S-ARIMA and the advantages of ARMA-SIN over S-ARIMA, just as intuitive understandings for readers. In Section \ref{sec:s-arima-insufficient}, we will explain the mathematical reason why S-ARIMA makes sense and point out its theoretical insufficiency. Finally, in Section \ref{sec:arma-sin}, the general ARMA-SIN methodology will be presented, and in Section \ref{sec:solution-to-warming-up}, we will analytically explain and derive the methods we presented in the warming-up Section \ref{sec:warming-up}. Glossaries of interest in this paper are placed in Appendix.

\begin{remark}\label{rem:arima-insufficiency}
When we mention the insufficiency of S-ARIMA, we actually mean its theoretical insufficiency instead of the prediction accuracy in some particular cases. This is because the exact model only outperforms other models when the problem is exact. For example, if the data is generated from a linear function with sufficiently small Gaussian white noise, then the linear regression model should be better than any other high-order polynomial regression model. However, we cannot claim that the linear model is best all the time. Note that the focused time series, namely the exact problem we study in this paper, is a wide-sense stationary stochastic process. Thus, the ARMA model, according to Wold's Decomposition theorem \cite{papoulis2002probability}, is the corresponding exact model, meaning the operator that makes the original time series exact as wide-sense stationary is better, because the ultimate issue is to train an ARMA model. It is in this sense that we assert the S-ARIMA model is insufficient.
\end{remark}

\section{Notations}\label{sec:notation}
Before we start, we should define some useful notations here.
\begin{enumerate}
  \item Let $\bm v = a:l:b$ define a vector $\bm v$ having the lower bound $a$, upper bound $b$ and step length $l$. For example, $\bm v = 0:0.1:0.5$ means $a=0$, $b=0.5$, and $l=0.1$. Thus $\bm v = [0, 0.1, 0.2, 0.3, 0.4, 0.5]^T$;
  \item Let the function $length({\bm x})$ return the length of the vector $\bm x$. For example, if $\bm x = [1,2,3]^T$, we have $length({\bm x}) = 3$;
  \item Let $\bm t$ denote the continuous time variable, and $\bm n$ its corresponding discrete time variable. For example, if $\bm t = 0:0.5:100$ (the time span is $100s$, and the sampling time is $T_s = 0.5s$), we will have $\bm n = \bm t/T_s = 0:1:length(t)-1 = 0:1:200$; let $N = length(\bm n)$;
  \item Let $\bm x$ denote a time series of interest;
  \item Let $randn(N)$ return a Gaussian white series (mean is zero, variance is one) with length of $N$;
  \item Let $ARMA(p,q|\bm {\varphi}, \bm {\theta})$ define an ARMA process with autoregressive order of $p$ and moving average order of $q$. Besides, the coefficient vectors $\bm {\varphi}$ and $\bm {\theta}$ are for autoregressive part and moving average part, respectively;
  \item Let the operator $\bm y = H(\bm x|\bm a, \bm b)$ define a difference equation (a.k.a., a linear system) as follows
    \begin{equation}\label{eq:ARMAPara}
        a_0 y_{k} + a_1 y_{k-1} +  a_2 y_{k-2} +...+ a_p y_{k-p} = b_0 x_{k} + b_1 x_{k-1} + b_2 x_{k-2} + ... + b_q x_{k-q},
    \end{equation}
  where $\bm a$, $\bm b$ are vectors and $length({\bm a}) = p + 1$, $length({\bm b}) = q + 1$.
\end{enumerate}

\section{Rationale Behind the S-ARIMA Method }\label{subsec:Wold's-Decomposition}
\begin{definition}[WSS Stochastic Process \cite{papoulis2002probability}]\label{def:WSS}
A real-valued stochastic process $x(t)$ is WSS if it satisfies:
\begin{itemize}
  \item \textbf{Invariant Mean}: $E\{x(t)\} = \eta$, where $\eta$ is a constant;
  \item \textbf{Invariant Autocorrelation}: $E\{x(t_1)x(t_2)\} = E\{x(t_1+\tau)x(t_1)\} = R(\tau)$, meaning it only depends on $\tau := t_2 - t_1$, having nothing to do with $t_1$.
\end{itemize}
\end{definition}
Invariant autocorrelation immediately admits the \textbf{Invariant Variance}, since $E\{x(t)\}^2 = R(0)$.

\begin{theorem}[Wold's Decomposition Theorem \cite{papoulis2002probability}]\label{thm:wold-decom}
Any WSS stochastic process $x(n)$ could be decomposed into two subprocess: (a) Regular process; and (b) Predictable process. Namely
\begin{equation}\label{eq:wold-decom}
  x(n) = x_r(n) + x_p(n),
\end{equation}
where $x_r(n)$ is a regular process and $x_p(n)$ is a predictable process. Furthermore, the two processes are orthogonal (implying uncorrelated): $E\{x_r(n+\tau)x_p(n)\} = 0$.
\end{theorem}

The detailed concepts of Regular Process (a.k.a. Rational-Spectra process from the spectral analysis perspective), and Predictable Process (a.k.a. Line-Spectra process) could be found in \cite{papoulis2002probability}. Intuitively, a {regular process} is mathematically as $x_r(n) = ARMA(p,q)$, and a {predictable process} is as $x_p(n) = \sum^m_{i=0} c_i e^{jw_in}$ for some non-negative integer $m$ where $w_i$ are non-random discrete frequencies \cite{diniz2010digital} with $w_0 = 0$, $E{c_i} = 0$, $E{c_i c^*_j} = 0$ when $i \ne j$, and $E{c^2_i} = \alpha_i > 0$. Note that the random coefficients $c_i$ of the complex-valued exponentials $e^{jw_in}$ has nothing to do with the time $n$. It means that $c_i$ are determined prior to $n=0$ [i.e., $c_i$ do not vary over time but they change in every different realization of $x(n)$]. Therefore, in one realization of $x(n)$, the predictable component $x_p(n)$ can be treated as a non-zero-mean but deterministic trend [which can be eliminated by the (seasonal) difference operator] of the regular component $x_r(n)$. This explains why the S-ARIMA method is rational for any WSS stochastic process: if this WSS stochastic process is regular, we directly use the ARMA model to fit; if it contains a predictable component, we alternatively use the S-ARIMA.

\section{Scenarios of Warming-up}\label{sec:warming-up}
As intuitive understandings for readers, we in this section provide some simulation scenarios to illustrate the theoretical insufficiency that S-ARIMA holds. Together with, the counterpart solutions given by ARMA-SIN will be also demonstrated. Following this warming-up, in the subsequent sections, we will progressively detail the motivations, philosophies, mathematics and methodologies of generating such useful ARMA-SIN solutions.

\subsection{ARMA Series of Ground Truth }
We generate a ARMA series for analysis later. Without loss of generality, we arbitrarily set
\begin{equation}\label{eq:ARMAExamplePara}
    \begin{array}{rl}
      \bm {\theta} &= [13, 5, 6]^T \\
      \bm {\varphi} &= [40, 2, 3, 6, 9]^T,
    \end{array}
\end{equation}
meaning the time series generated from a Gaussian white series $\epsilon_k$ is given as
\begin{equation}\label{eq:ARMAExampleTruth}
    40x^0_{k} + 2x^0_{k-1} + 3x^0_{k-2} + 6x^0_{k-3} + 9x^0_{k-4} = 13{\epsilon}_{k} + 5{\epsilon}_{k-1} + 6{\epsilon}_{k-2},
\end{equation}
where $k$ denotes the discrete time index (namely $k \in \bm n$) and $x^0_k := 0, \epsilon_k := 0$ if $k < 0$. However, this discrete linear system is guaranteed to be minimum-phase stable \cite{diniz2010digital}. That is, it is stable and inversely stable.

Therefore, if the focused raw time series is as $\bm x = f({\bm x}^0)$ (for example $\bm x = {\bm x}^0 + \bm n$, i.e., linear trend), the operator that exactly transforms $\bm x$ to its wide-sense stationary component ${\bm x}^0$ should be the best. This is because we finally aim to use the ARMA model to fit the transformed series. Let ${\bmh x}^0$ be the transformed series from $\bm x$. The nearer between ${\bmh x}^0$ and ${\bm x}^0$, the better. For more, see Remark \ref{rem:arima-insufficiency}.

\subsection{The Case of Variant Mean}
In this subsection, we investigate a scenario being with variant mean, implying it is non-stationary in the mean sense. Let $\bm t = 0:0.5:100$ (viz., $T_s = 0.5$), $\bm n = \bm t/T_s$, and $\bm x = {\bm x}^0 + 0.1\bm t$. It means the trend component is a linear function.

\begin{figure}[htbp]
    \centering
    \subfigure[Transformed series]{
        \begin{minipage}[htbp]{0.45\linewidth}
            \centering
            \includegraphics[height=4cm]{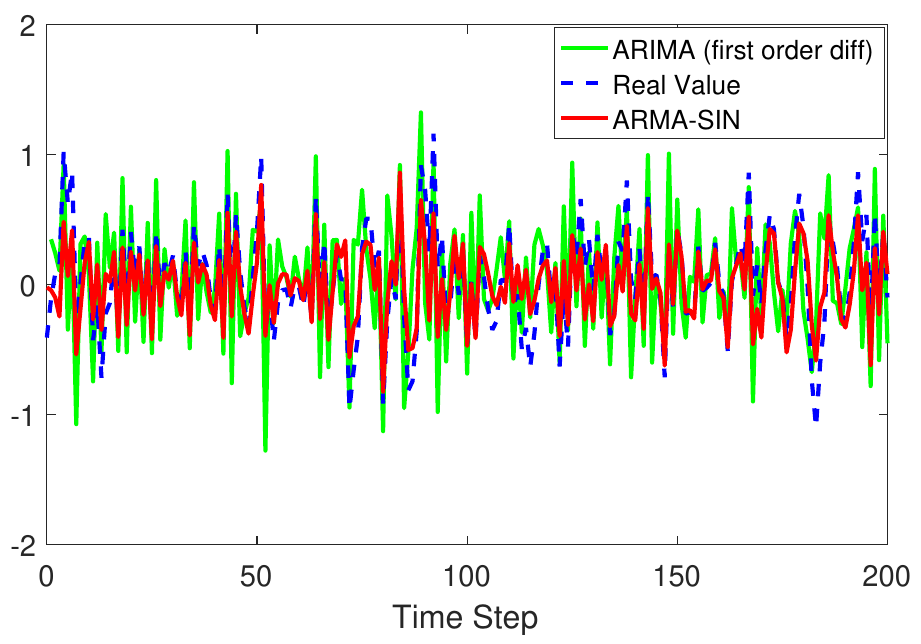}
        \end{minipage}
    }
    \subfigure[Transformed series (locally enlarged)]{
        \begin{minipage}[htbp]{0.45\linewidth}
            \centering
            \includegraphics[height=4cm]{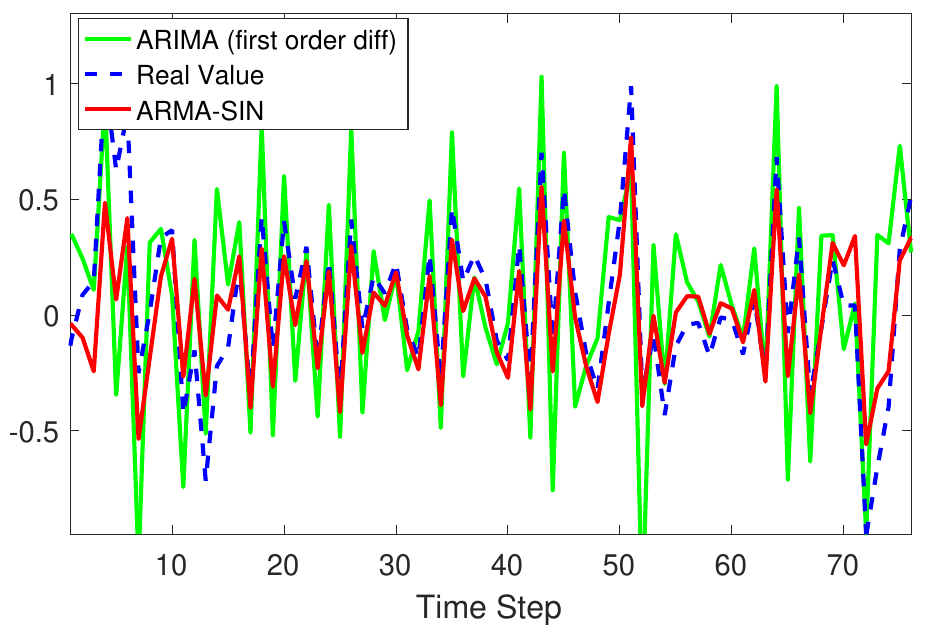}
        \end{minipage}
    }

    \subfigure[Prediction results]{
        \begin{minipage}[htbp]{0.45\linewidth}
            \centering
            \includegraphics[height=4cm]{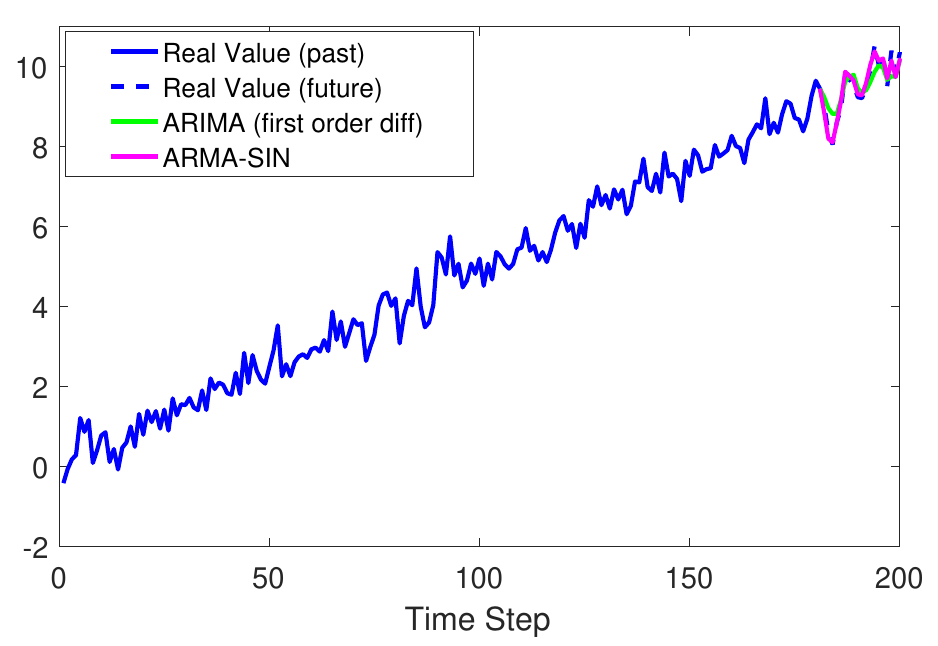}
        \end{minipage}
    }
    \subfigure[Prediction results (locally enlarged)]{
        \begin{minipage}[htbp]{0.45\linewidth}
            \centering
            \includegraphics[height=4cm]{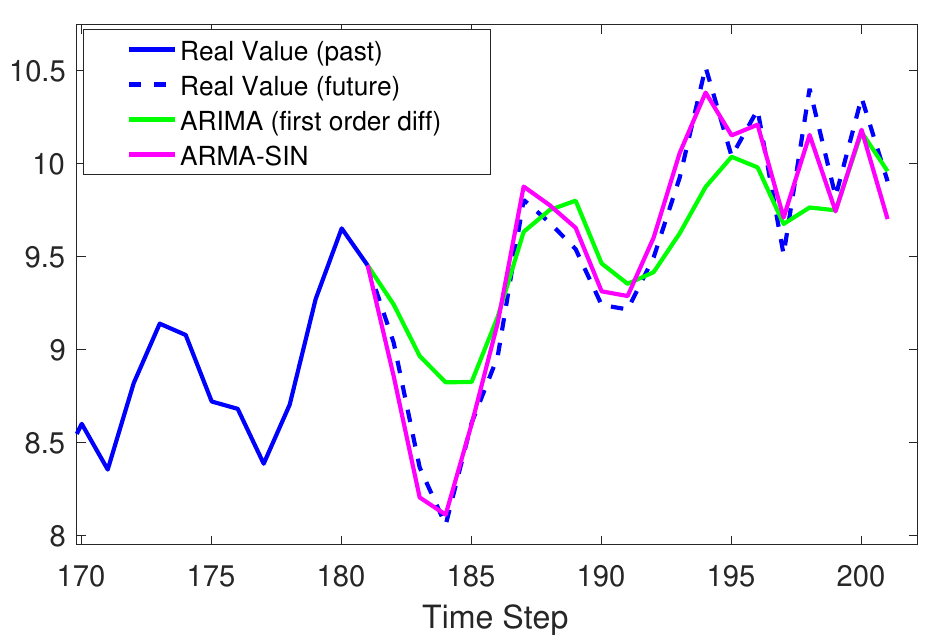}
        \end{minipage}
    }

    \centering
    \caption{Transformed series and prediction results for variant mean (diff: difference).}
    \label{fig:warmingup-case1}
\end{figure}

If we follow the standard modelling procedure with Box-Jenkins (ARIMA) method \cite{hyndman2018forecasting,calheiros2015workload,box2015time,hamilton1995time}, we have the estimated ARIMA(4,1,2) model to handle this problem, meaning the operator used is the first order differencing. Instead, if we use ARMA-SIN method, we have
\begin{equation}\label{eq:warmingup-case1}
  {\bmh x}^0 = H(\bm x|\bm a,\bm b),
\end{equation}
where $$\bm a = [1, -1.7101, 1.3712, -0.3152]^T,$$ $$\bm b = [0.6226, -1.5757, 1.5757, -0.6226]^T,$$ and the ARMA part for approximating ${\bm x}^0$ is ARMA(4,1).  The results of two methods are shown in Figure \ref{fig:warmingup-case1}.

Besides, we have $100$ times of monte carlo simulation and the averaged prediction MSE is given in Table \ref{tab:warmingup-case1}.  Clearly, the transformed series ${\bmh x}^0$ of ARMA-SIN is nearer to its ground truth ${\bm x}^0$ than that of ARIMA. Thus the prediction accuracy is more satisfactory.
\begin{table}[htbp]
	\centering
	\caption{Averaged prediction MSE of ARIMA and ARMA-SIN for variant mean}
	\label{tab:warmingup-case1}
	\begin{tabular}{lcccc}
		\toprule
		& ARIMA     & ARMA-SIN \\
		\hline
		MSE   & 0.0977    &  0.0203 \\
		\bottomrule
	\end{tabular}
\end{table}

\subsection{The Case of Using SARIMA}\label{sebsec:case-3}
In this subsection, we investigate a scenario that is suitable for using SARIMA. Let $\bm t = 0:0.1:50$ (viz., $T_s = 0.1$), $\bm n = \bm t/T_s$, and $\bm x = {\bm x}^0 + sin(5\bm t)$.  It means the trend component is a sine function.

\begin{figure}[htbp]
    \centering
    \subfigure[Transformed series]{
        \begin{minipage}[htbp]{0.45\linewidth}
            \centering
            \includegraphics[height=4cm]{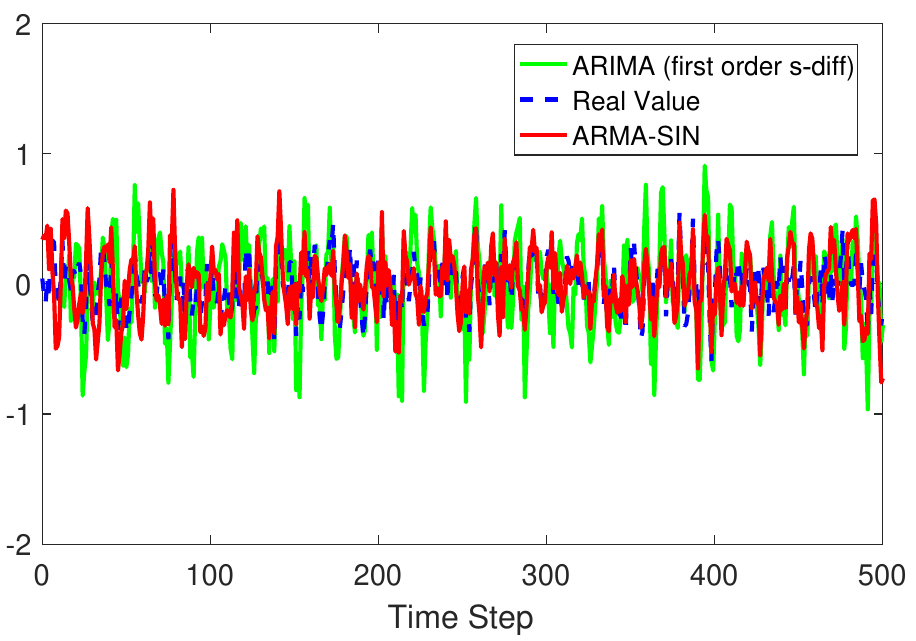}
        \end{minipage}
    }
    \subfigure[Transformed series (locally enlarged)]{
        \begin{minipage}[htbp]{0.45\linewidth}
            \centering
            \includegraphics[height=4cm]{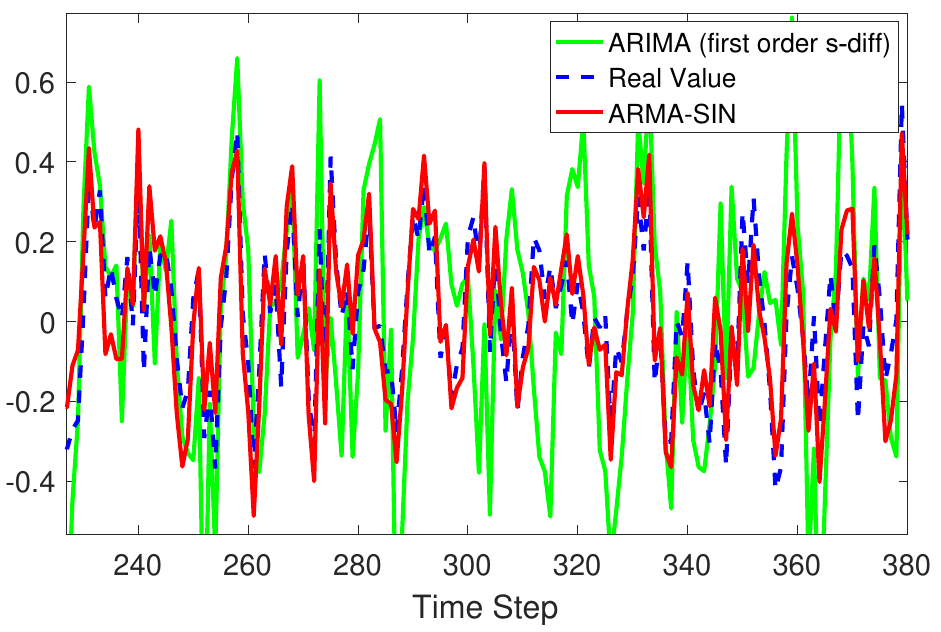}
        \end{minipage}
    }

    \subfigure[Prediction results]{
        \begin{minipage}[htbp]{0.45\linewidth}
            \centering
            \includegraphics[height=4cm]{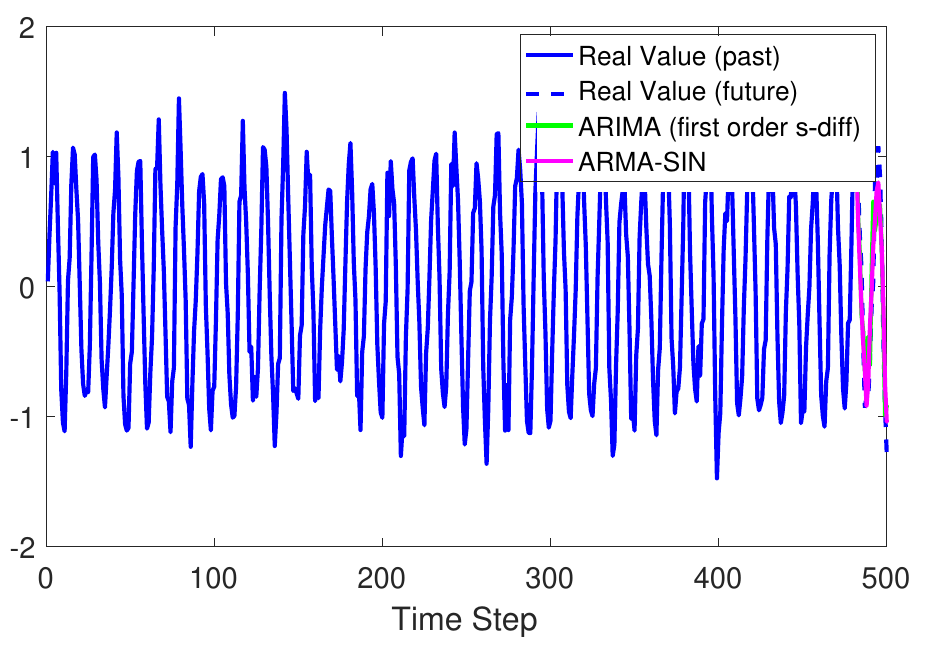}
        \end{minipage}
    }
    \subfigure[Prediction results (locally enlarged)]{
        \begin{minipage}[htbp]{0.45\linewidth}
            \centering
            \includegraphics[height=4cm]{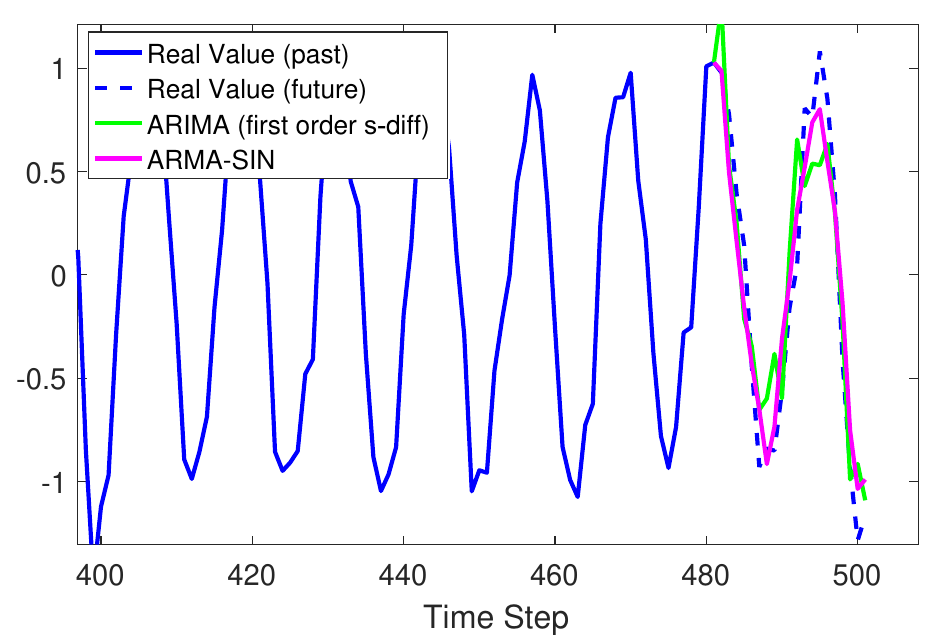}
        \end{minipage}
    }

    \centering
    \caption{Transformed series and prediction results of SARIMA and ARMA-SIN (s-diff: seasonal difference).}
    \label{fig:warmingup-case3}
\end{figure}

If we follow the standard modelling procedure with Box-Jenkins (ARIMA) method, we have the estimated SARIMA(4,1,1)(12,1,0,0) model to handle this problem, meaning the operator used is the $12$-lag seasonal differencing. Instead, if we use ARMA-SIN method, we have
\begin{equation}\label{eq:warmingup-case3}
  {\bmh x}^0 = H(\bm x|\bm a,\bm b),
\end{equation}
where $$\bm a = [1, -5.1801, 11.8864, -15.30778, 11.6563, -4.9815, 0.9430]^T,$$ $$\bm b = [0.9713, -5.0804, 11.7716, -15.3086, 11.7716, -5.0804, 0.9713]^T,$$ and the ARMA part for approximating ${\bm x}^0$ is ARMA(4,1).  The results of two methods are shown in Figure \ref{fig:warmingup-case3}.

Besides, we have $100$ times of monte carlo simulation and the averaged prediction MSE is given in Table \ref{tab:warmingup-case3}.
\begin{table}[htbp]
    \centering
    \caption{Averaged prediction MSE of SARIMA and ARMA-SIN}
    \label{tab:warmingup-case3}
    \begin{tabular}{lcccc}
        \toprule
                & ARIMA     & ARMA-SIN \\
        \hline
          MSE   & 0.1022    &  0.0516 \\
        \bottomrule
    \end{tabular}
\end{table}
Clearly, the transformed series ${\bmh x}^0$ of ARMA-SIN is nearer to its ground truth ${\bm x}^0$ than that of SARIMA. Thus the prediction accuracy is more satisfactory.

\subsection{The Case of Directly Estimating the Seasonal Component}
In this subsection, we investigate a scenario that we can directly estimate the seasonal component. Let $\bm t = 0:0.1:10$ (viz., $T_s = 0.1$), $\bm n = \bm t/T_s$, and $\bm x = {\bm x}^0 + f(\bm t)$, where $f(\bm t) = sin(2\bm t)$. Our purpose is to estimate out the function $sin(2t)$ directly from the collected history data.

If we use our ARMA-SIN method, we can know that $f(x)$ is with the form as
\begin{equation}\label{eq:warmingup-case4}
    \begin{array}{cll}
        {\hat f}(t) & = 0.9721 cos(0.2001 n - 1.5440)\\
                    & = 0.9721 cos(0.2001 t/T_s - 1.5440)\\
                    & = 0.9721 cos(2.001 t - 0.4945 \pi)
    \end{array}
\end{equation}
It is amazingly close to its ground truth of $f(t) = sin(2t) = cos(2t - \pi/2) = cos(2t - 0.5\pi)$.

If we follow the standard modelling procedure with Box-Jenkins methodology, we have the estimated SARIMA(4,0,1)(31,0,0,0) model to handle this problem, meaning the operator used is the $31$-lag seasonal differencing.

\begin{figure}[htbp]
    \centering
    \subfigure[Transformed series]{
        \begin{minipage}[htbp]{0.45\linewidth}
            \centering
            \includegraphics[height=4cm]{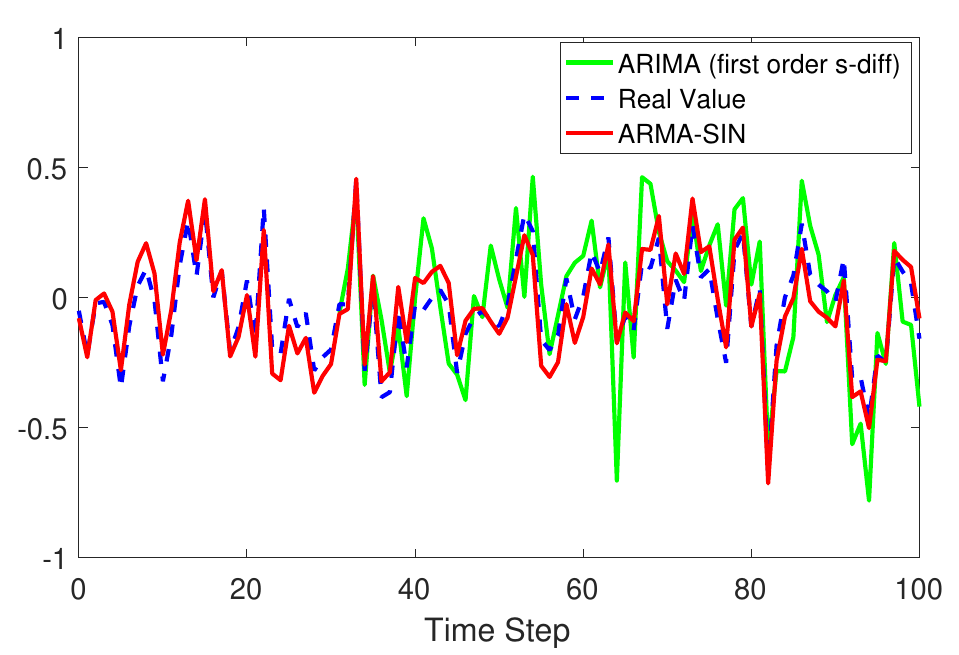}
        \end{minipage}
    }
    \subfigure[Transformed series (locally enlarged)]{
        \begin{minipage}[htbp]{0.45\linewidth}
            \centering
            \includegraphics[height=4cm]{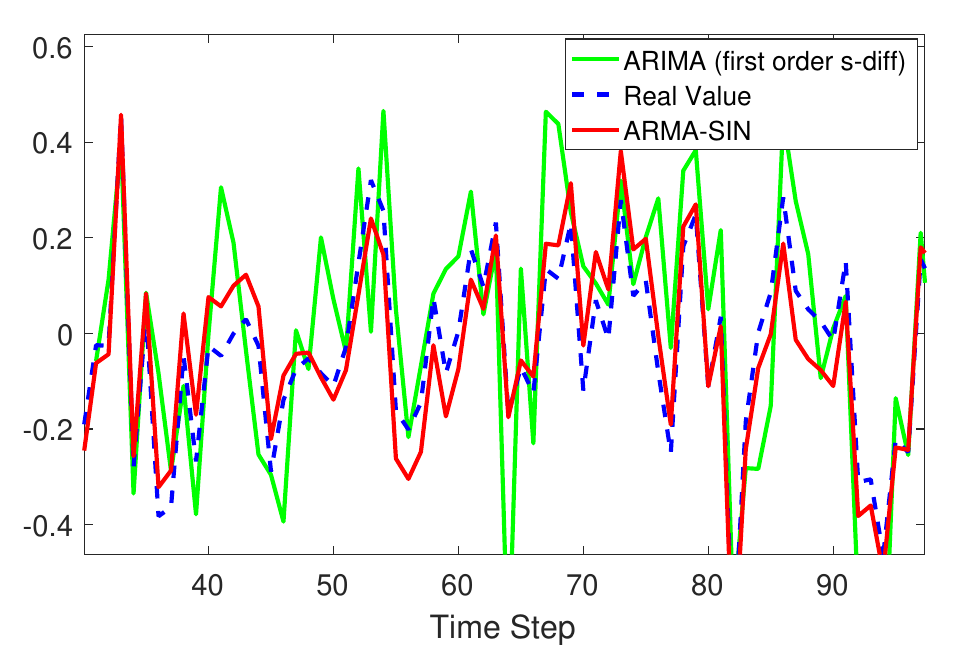}
        \end{minipage}
    }

    \subfigure[Prediction results]{
        \begin{minipage}[htbp]{0.45\linewidth}
            \centering
            \includegraphics[height=4cm]{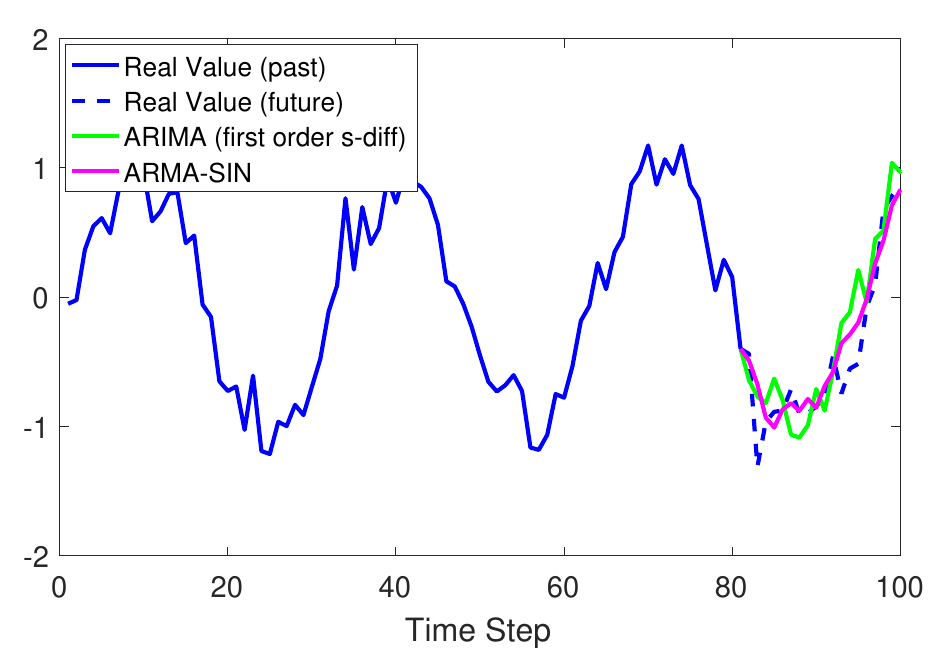}
        \end{minipage}
    }
    \subfigure[Prediction results (locally enlarged)]{
        \begin{minipage}[htbp]{0.45\linewidth}
            \centering
            \includegraphics[height=4cm]{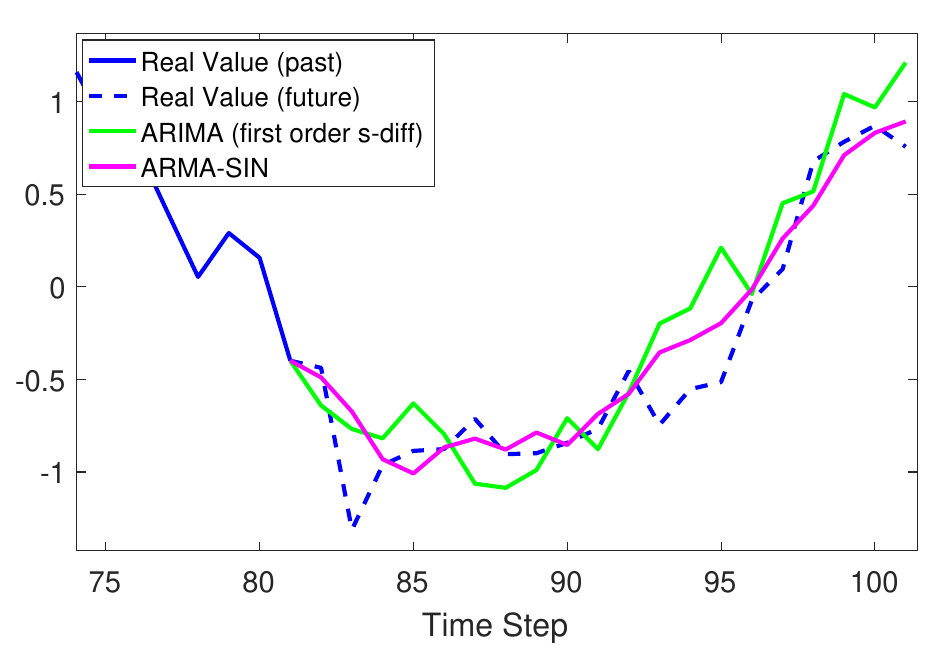}
        \end{minipage}
    }

    \centering
    \caption{Transformed series and prediction results of direct estimation.}
    \label{fig:warmingup-case4}
\end{figure}

The results of two methods are shown in Figure \ref{fig:warmingup-case4}.

Besides, we have $100$ times of monte carlo simulation and the averaged prediction MSE is given in Table \ref{tab:warmingup-case4}.

\begin{table}[htbp]
    \centering
    \caption{Averaged prediction MSE of direct estimation}
    \label{tab:warmingup-case4}
    \begin{tabular}{lcccc}
        \toprule
                & SARIMA     & ARMA-SIN \\
        \hline
          MSE   & 0.0844    &   0.0355 \\
        \bottomrule
    \end{tabular}
\end{table}

Clearly, the transformed series ${\bmh x}^0$ of ARMA-SIN is nearer to its ground truth ${\bm x}^0$ than that of SARIMA. Thus, the prediction accuracy is more satisfactory.

Intuitively, we can see from the three examples above that the ARMA-SIN method is indeed interesting over the S-ARIMA method. In the following sections, we will progressively explain the philosophy behind and detail the ARMA-SIN method.

\section{Secret Behind the ARIMA and SARIMA}\label{sec:s-arima-insufficient}
\subsection{Nature of Difference Operator and Seasonal Difference Operator}\label{sec:s-arima-make-sense}
As mentioned in the previous section, the ARIMA and SARIMA attempt to make stationary a stochastic process by difference operator and seasonal difference operator. In this section we will investigate the nature of ARIMA and SARIMA from the perspective of SADFA (see Appendix). Specifically, we should pay our attention to the nature of the difference operator and seasonal difference operator.

\begin{theorem}\label{thm:ARIMA}
The nature of $d$-order difference operator is actually a high-pass digital filter that denies the low-frequency components of a time series. Since the trend of a time series is generally the low-frequency components (see Discrete Fourier Transform \cite{diniz2010digital}), the $d$-order difference operator makes sense to make stationary a non-stationary (in the mean sense) stochastic process.
\end{theorem}
\begin{proof}
The transfer function of the $d$-order difference operator is given as
\begin{equation}\label{eq:Hz-d-diff-operator}
  H(z) = (1 - z^{-1})^d,
\end{equation}
and the amplitude-frequency response is as
\begin{equation}\label{eq:Hjw-d-diff-operator}
  |H(e^{jw})| = |(1 - e^{-jw})^d| = [\sqrt{2-2\cos(w)}]^d.
\end{equation}
Eq. (\ref{eq:Hjw-d-diff-operator}) immediately admits the theorem, since in the interval $[0,~\pi]$,  $|H(e^{jw})|$ is increasing from zero. Intuitively, see Figure \ref{fig:d-diff-operator}.
\begin{figure}[htbp]
    \centering
    \subfigure[1-order difference operator]{
        \begin{minipage}[htbp]{0.45\linewidth}
            \centering
            \includegraphics[height=4cm]{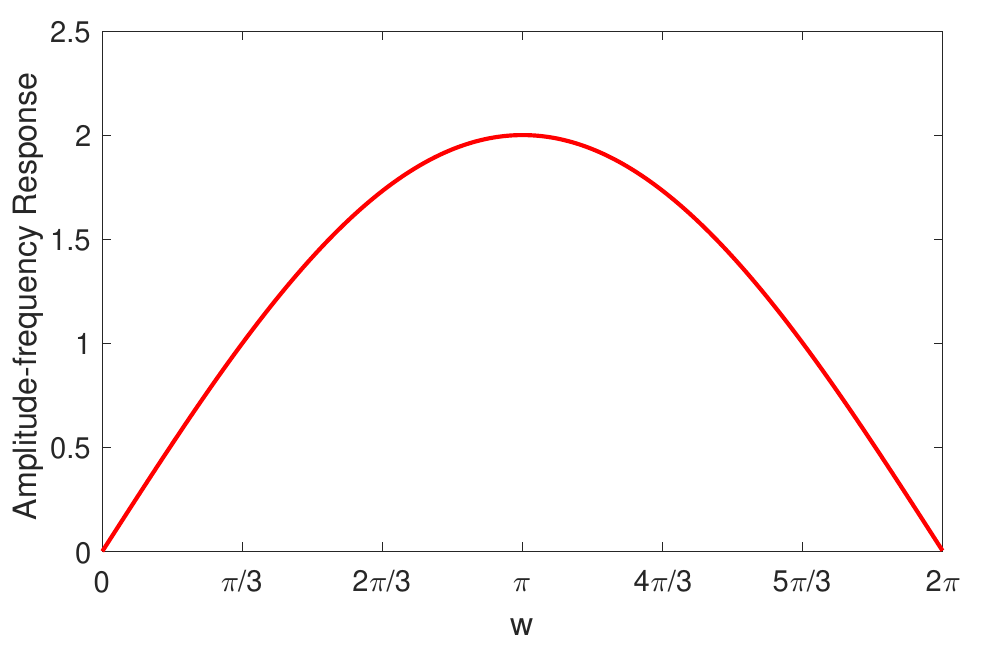}
        \end{minipage}
    }
    \subfigure[4-order difference operator]{
        \begin{minipage}[htbp]{0.45\linewidth}
            \centering
            \includegraphics[height=4cm]{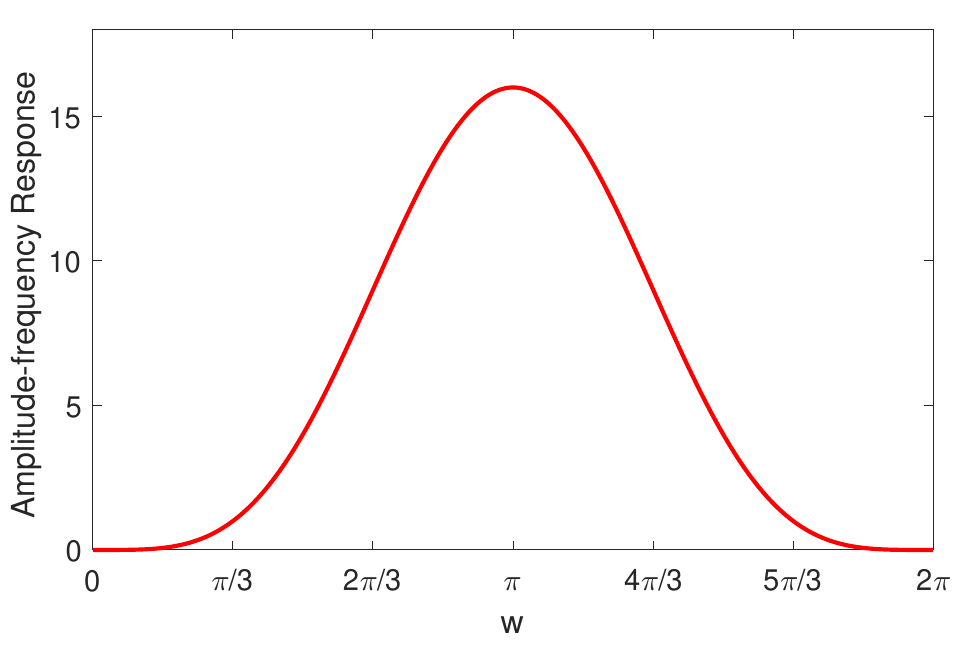}
        \end{minipage}
    }
    \centering
    \caption{Amplitude-frequency responses of difference operator.}
    \label{fig:d-diff-operator}
\end{figure}
\end{proof}

\begin{theorem}\label{thm:SARIMA}
The nature of $L$-lag seasonal difference operator (L-SDO) is actually a comb digital filter that denies some frequency components with $w = 2k\pi/L, k = 0,1,2,...,L-1$. Since the seasonal components (namely periodic components) with period $L$ has period $L$ in its spectra as well (meaning the frequencies of seasonal components are $w = 2k\pi/L, k = 0,1,2,...,L-1$, see Discrete Fourier Series \cite{diniz2010digital}) the $L$-lag difference operator makes sense to make stationary a non-stationary (in the mean sense) stochastic process.
\end{theorem}
\begin{proof}
The transfer function of the $L$-lag seasonal difference operator is given as
\begin{equation}\label{eq:Hz-L-lag-operator}
  H(z) = 1 - z^{-L},
\end{equation}
and the amplitude-frequency response is as
\begin{equation}\label{eq:Hjw-L-lag-operator}
    \begin{array}{cll}
        |H(e^{jw})| &= |1 - e^{-jLw}| \\
                    &= |[1-\cos(Lw)] + j \sin(Lw)| \\
                    &= \sqrt{2 - 2\cos(Lw)}.
    \end{array}
\end{equation}
Eq. (\ref{eq:Hjw-L-lag-operator}) immediately admits the theorem, since in the interval $[0,~\pi]$,  $|H(e^{jw})|$ are zero-valued at $w = 2k\pi/L, k = 0,1,2,3,...,L-1$. Intuitively, we consider a sine series $x(n)$ with period $L=12$ ($t=0:2\pi/L:100$) and take $12$-lag seasonal difference over it. The results are illustrated in Figure \ref{fig:L-lag-operator}.
\begin{figure}[htbp]
    \centering
    \subfigure[Full picture]{
        \begin{minipage}[htbp]{0.45\linewidth}
            \centering
            \includegraphics[height=4cm]{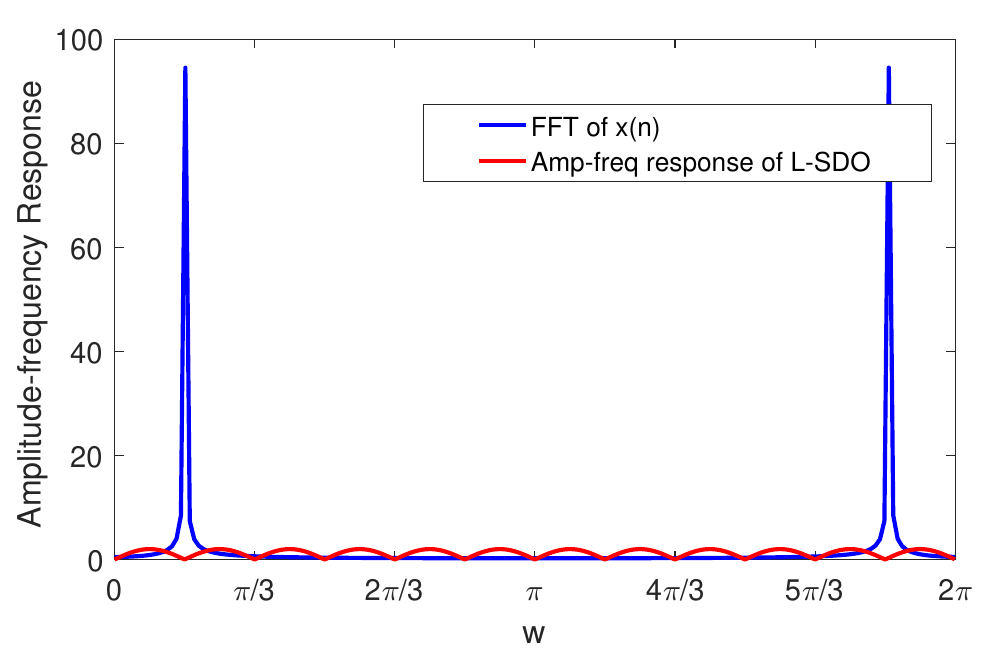}
        \end{minipage}
    }
    \subfigure[Locally enlarged]{
        \begin{minipage}[htbp]{0.45\linewidth}
            \centering
            \includegraphics[height=4cm]{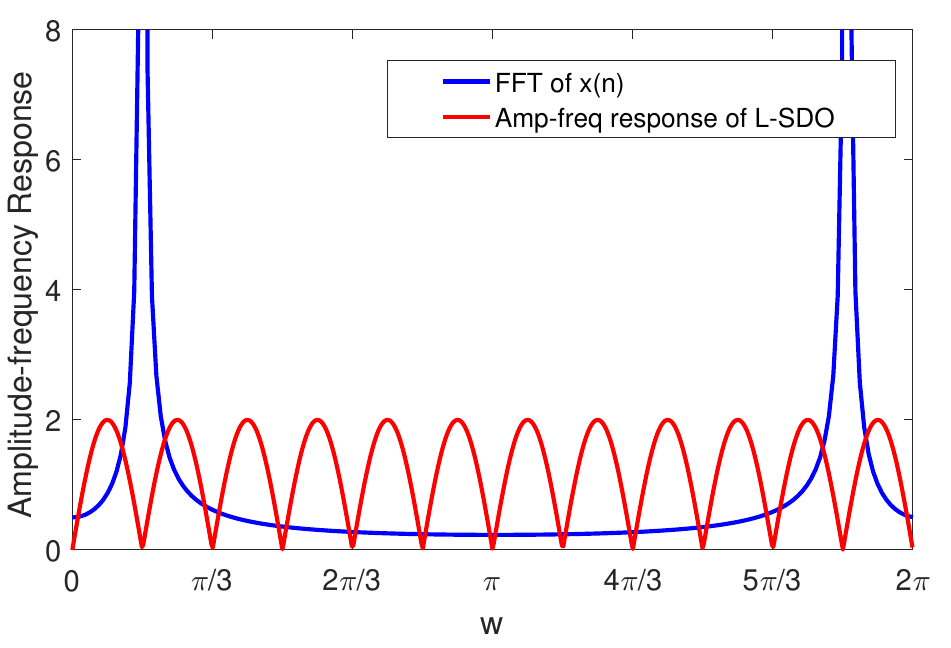}
        \end{minipage}
    }
    \centering
    \caption{FFT of $x(n)$ and Amplitude-frequency responses of $L$-lag seasonal difference operator.}
    \label{fig:L-lag-operator}
\end{figure}
\end{proof}

\begin{remark}
Recall that in discrete Fourier frequency domain, every $|H(e^{jw})|$ is periodic over $w$ and the period is $2\pi$. Since the highest frequency is indicated by $w = \pi$ and lowest is by $w=0$ and $w = 2\pi$, we should only pay attention to the interval $[0,~2\pi]$. The trend of a time series is usually in low-frequency interval (around $w=0$ and $w=2 \pi$) with large values (e.g., see Figure \ref{fig:case1-detail} (a), $w$ below $\pi/6$). Besides, according to Discrete Fourier Series \cite{diniz2010digital}, the seasonal component has impulses uniformly distributed (uniformly spaced) in the all frequency domain $w \in [0,~\pi]$ (e.g., see Figure \ref{fig:case3-detail} (a), around $w=\pi/6$). Therefore, in frequency domain, we expect to remove all these outliers so that the transformed time series would become stationary.
\end{remark}

\begin{remark}
Note that if the seasonal component of a time series is perfectly a sine function, it only has one impulse in its spectra because the other impulses are zero-valued. For more, see Discrete Fourier Series \cite{diniz2010digital}. This is why Figure \ref{fig:case3-detail} (a) only has one impulse other than many.
\end{remark}

\subsection{Why Are the ARIMA and SARIMA Not Sufficient}\label{sec:s-arima-insufficient-subsec}
By now, Theorem \ref{thm:ARIMA} and Theorem \ref{thm:SARIMA} support the effectiveness of S-ARIMA somewhere. However, it is obvious that except the unwanted frequencies, $d$-order difference operator of ARIMA and $L$-lag seasonal difference operator of SARIMA also negatively impact the desired frequencies which should remain unchanged, for example, the frequencies higher than $w=\pi/3$ in Figure \ref{fig:d-diff-operator} have been significantly and unwantedly amplified, and the frequencies at and around $w = 2k\pi/12, k = 0,2,3,...11, \text{(i.e.}, k \neq 1)$ in Figure \ref{fig:L-lag-operator} have also been unwantedly wiped away. This raises the insufficiency of S-ARIMA. Thus we have Theorem \ref{thm:insufficiency-S-ARIMA}.

\begin{theorem}\label{thm:insufficiency-S-ARIMA}
The ARIMA and SARIMA model are theoretically insufficient since they also negatively impact (distort or eliminate) the desired frequency components (points or intervals).
\end{theorem}
\begin{proof}
Due to Theorem \ref{thm:ARIMA}, Theorem \ref{thm:SARIMA} and the statement itself in this theorem, the conclusion stands.
\end{proof}

\begin{remark}
Although the ARIMA and SARIMA are theoretically insufficient, they are easiest methods to handle the non-stationary problems in practice. It means instead of designing some more proper (but more complicated) operators to make stationary a time series, we can for simplicity choose (seasonal) difference operator, if the performances are satisfactory for some specific problems in engineering. It is simplicity of such difference operators that makes them popular in engineering over years.
\end{remark}

\section{The General ARMA-SIN Model}\label{sec:arma-sin}
According to Theorem \ref{thm:wold-decom}, any WSS time series has two components: an ARMA part and a complex-valued exponential sum (i.e., sum of sine time functions) part. Therefore, the ARMA-SIN \textbf{model} is for a general non-stationary time series where SIN means the sum of SINe functions. Besides, our analysis in this paper is based on SADFA (spectral analysis and digital filtering approach) and Box-Jenkins methodology. In detail, when we have a general time series, we first use SADFA to transform it to be wide-sense stationary and then use Box-Jenkins methodology to model the transformed series. Thus, the term ARMA-SIN in this paper is also referred to a time series analysis \textbf{method} where SIN is for SADFA and ARMA is for Box-Jenkins methodology.

In Algorithm \ref{alg:ARMA-SIN}, we detail the methodology of ARMA-SIN.

\begin{algorithm}[htbp]
	\caption{ARMA-SIN}
	\label{alg:ARMA-SIN}
	\begin{algorithmic}[1]
        \State \textbf{Spectral Analysis:} transform the focused time series into Fourier frequency domain and identify the impulses frequency points (or interval) in frequency domain. Usually, points for the seasonal components and interval (low-frequency region around $w=0$) for the trend;
        \State \textbf{Digital Filtering:} design a proper digital filter with proper cut-off frequencies such that the focused time series could be made stationary. Ideally, such proper digital filter should only remove the unwanted frequency points or interval while keeping the rest unchanged;
        \State \textbf{ARMA:} follow the standard Box-Jenkins methodology to model the stationary remainder (transformed time series) as a ARMA one;
        \State \textbf{Forecast:} predict the future with two independent process, ARMA part and the trend (and/or seasonal) part, respectively, and then integrate the results together. Note that the raw time series subtract the ARMA remainder gives the trend (or seasonal) part.
	\end{algorithmic}
\end{algorithm}

\begin{remark}
As we can see, the ARIMA and SARIMA are special cases of Algorithm \ref{alg:ARMA-SIN}. For ARIMA, the digital filter we use is just the difference operator (i.e., a special case of high-frequency-pass filters, that is, a low-frequency-stop filter) to remove (stop) the low-frequency trend part. For SARIMA, the digital filter we use is just the seasonal difference operator (i.e., a special case of comb filters, that is, a fixed-point-frequency-stop filter) to remove (stop) the frequency impulses in frequency domain.
\end{remark}

\section{Derive the Solutions for Scenarios of Warming-up}\label{sec:solution-to-warming-up}
In this section, we will detail the derivation for solutions of scenarios of warming-up given in Section \ref{sec:warming-up}. For briefness, we will not repeat the standard modelling procedures of Box-Jenkins methodology \cite{box2015time}. 
All the filters this paper designed are based on IIR (infinite impulse response) model and Elliptic method \cite{diniz2010digital}. 

\begin{remark}\label{rem:ARMA-SIN-parameters}
Usually it is hard to decide the best parameters like cut-off frequencies at the first glance for a general problem. However, we could try some times with plausible parameters and choose the best among them. This seems tedious but unavoidable in practice because there is no free lunch, advanced method means complex coding, complex parameters selecting and/or complex calculation burden. In this sense, the use of simple (seasonal) difference operator seems really delightful for all of us. Nevertheless, the trade-off is the unsatisfactory performances and the powerlessness somewhere.
\end{remark}

\subsection{The Case of Variant Mean}
The time series is shown in Figure \ref{fig:warmingup-case1} (c). The trend is linear. According to Discrete Fourier Transform theory, the linear trend could be represented by a finite sum of sine functions with acceptable approximation error. In order to separate (extract) the linear trend, we should design a high-pass filter which denies the low-frequency component (trend) to pass and allows the high-frequency components to go through without any changes. Since we use the Elliptic method to design a IIR filter (the function \textit{ellipord} in MATLAB, for more, see its reference page in \cite{ellipord}), we set the parameters in Table \ref{tab:warmingup-case1-parameters}.
\begin{table}[htbp]
    \centering
    \caption{The filter parameters for the case of variant mean}
    \label{tab:warmingup-case1-parameters}
    \begin{tabular}{lcccc}
        \toprule
                    & Wp        & Ws        & Rp    & Rs \\
        \hline
          Value     & 0.25    &  0.2        & 1     & 10 \\
        \bottomrule
    \end{tabular}
\end{table}

The specific meanings and detailed usage of parameters in Table \ref{tab:warmingup-case1-parameters} should be found in \cite{diniz2010digital} or the reference page of the function \textit{ellipord} in \cite{ellipord}.

This high-pass filter actually defines the system (\ref{eq:warmingup-case1}).

\begin{figure}[htbp]
    \centering
    \subfigure[Spectrum of $x(n)$]{
        \begin{minipage}[htbp]{0.45\linewidth}
            \centering
            \includegraphics[height=4cm]{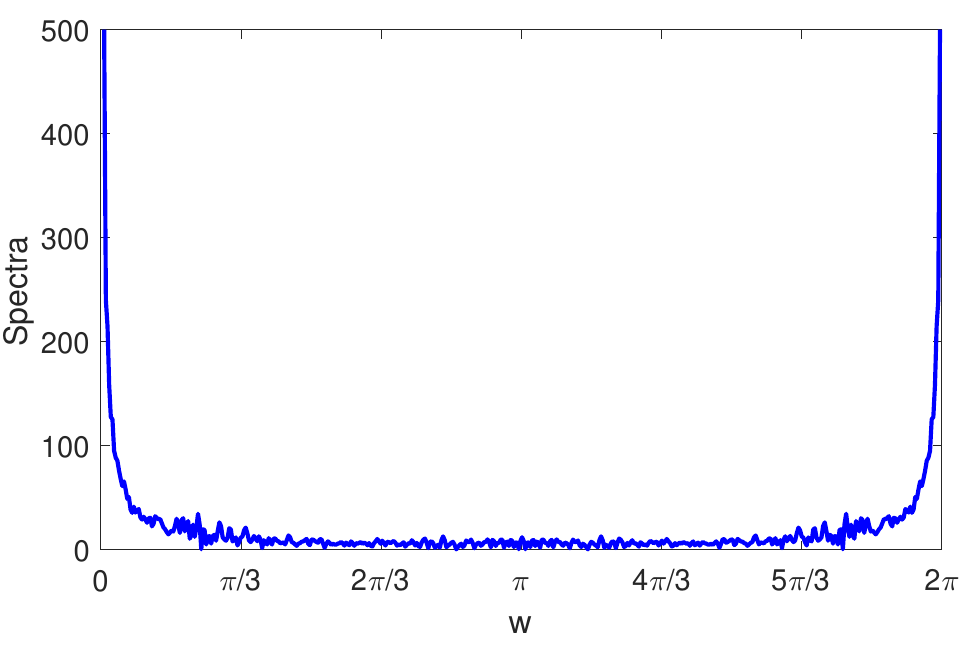}
        \end{minipage}
    }
    \subfigure[Digital filter to detrend the $x(n)$]{
        \begin{minipage}[htbp]{0.45\linewidth}
            \centering
            \includegraphics[height=4cm]{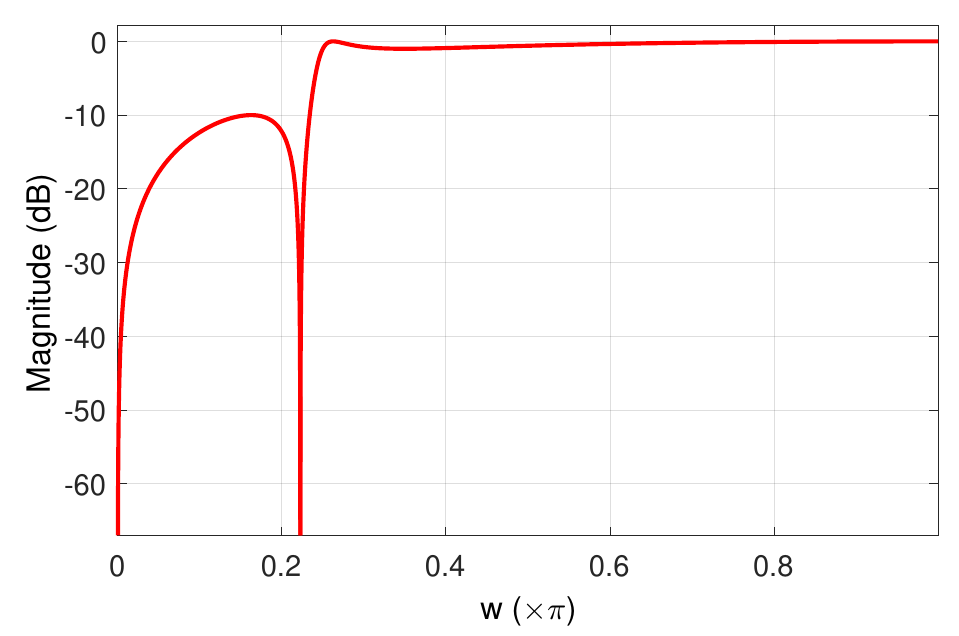}
        \end{minipage}
    }

    \subfigure[Estimated spectra of $x^0$ using ARMA-SIN]{
        \begin{minipage}[htbp]{0.45\linewidth}
            \centering
            \includegraphics[height=4cm]{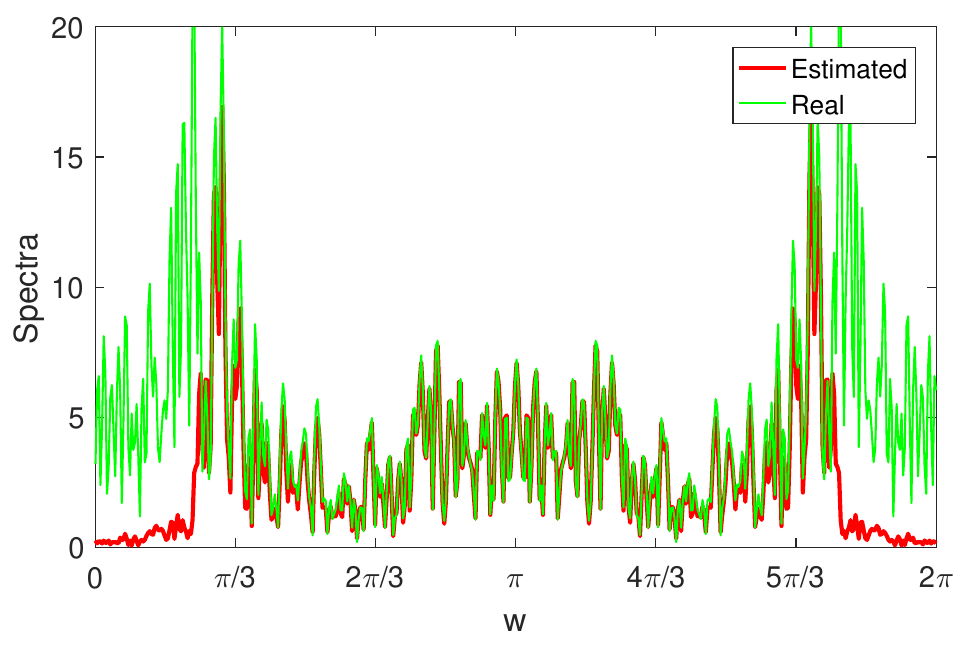}
        \end{minipage}
    }
    \subfigure[Estimated spectra of $x^0$ using $1$-order difference]{
        \begin{minipage}[htbp]{0.45\linewidth}
            \centering
            \includegraphics[height=4cm]{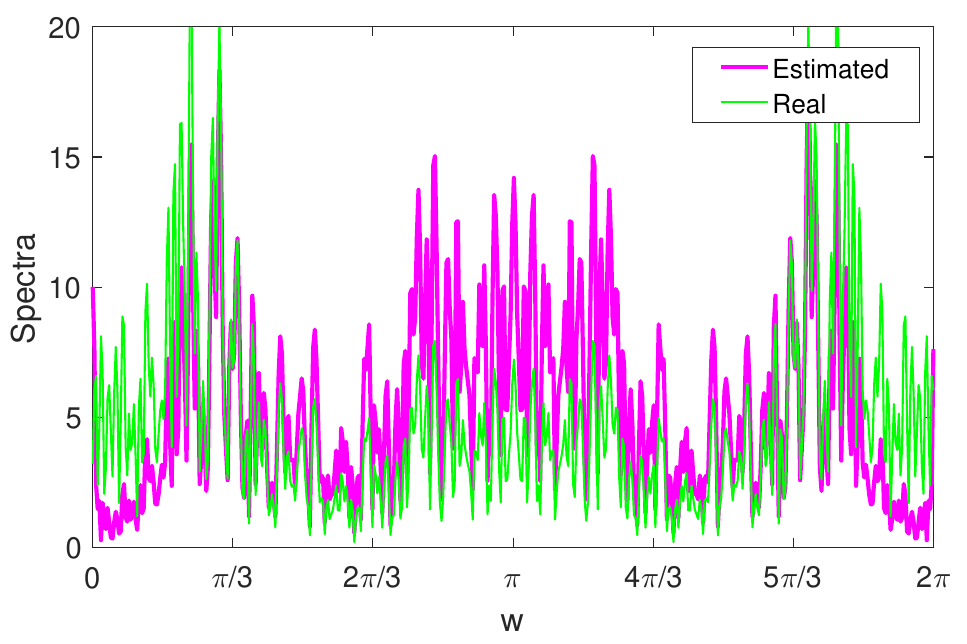}
        \end{minipage}
    }
    \centering
    \caption{Spectra analysis for the case of variant mean.}
    \label{fig:case1-detail}
\end{figure}

The spectral analysis results are given in Figure \ref{fig:case1-detail}.

From Figure \ref{fig:case1-detail}, we can see the $1$-order difference operator significantly distorts (amplifies) the spectra in high frequency interval, although it is powerful to detrend. However, our ARMA-SIN keeps $1$ [n.b., $20\log(1) = 0dB$] in higher frequency area, making no changes to desired components. This is why our ARMA-SIN outperforms ARIMA. Note that Figure \ref{fig:case1-detail} (b) is given in logarithmic unit of decibel ($dB$). For more on decibel, see \cite{diniz2010digital}.

\subsection{The Case of Using SARIMA}
\begin{figure}[htbp]
    \centering
    \subfigure[Spectrum of $x(n)$]{
        \begin{minipage}[htbp]{0.45\linewidth}
            \centering
            \includegraphics[height=4cm]{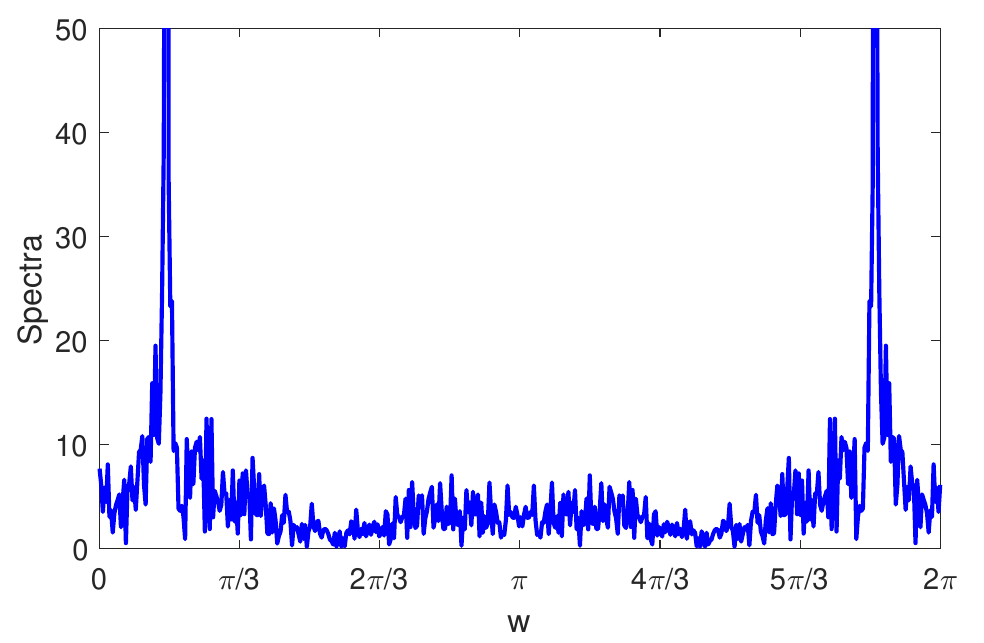}
        \end{minipage}
    }
    \subfigure[Digital filter to detrend the $x(n)$]{
        \begin{minipage}[htbp]{0.45\linewidth}
            \centering
            \includegraphics[height=4cm]{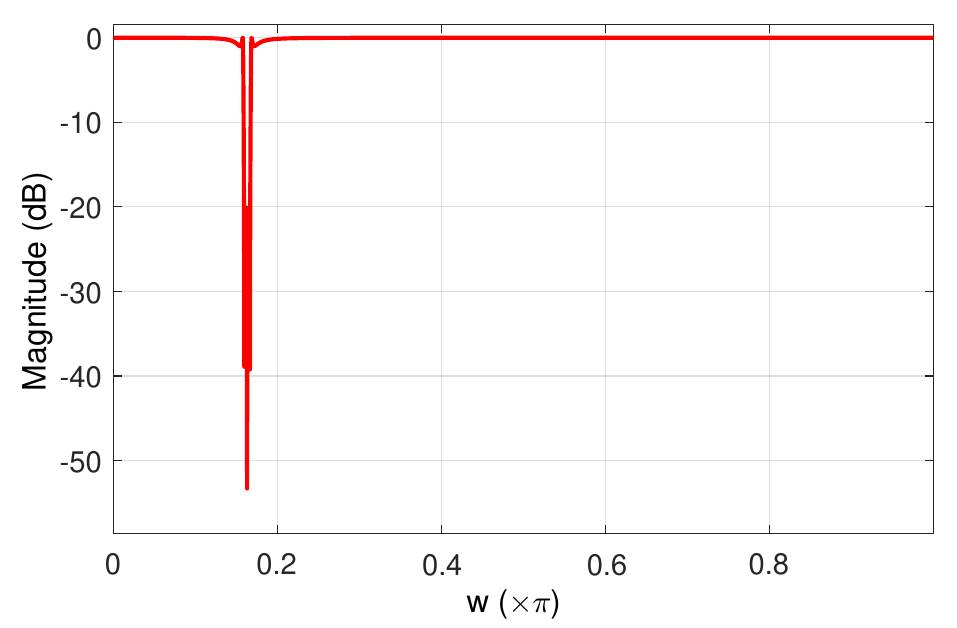}
        \end{minipage}
    }

    \subfigure[Estimated spectra of $x^0$ using ARMA-SIN]{
        \begin{minipage}[htbp]{0.45\linewidth}
            \centering
            \includegraphics[height=4cm]{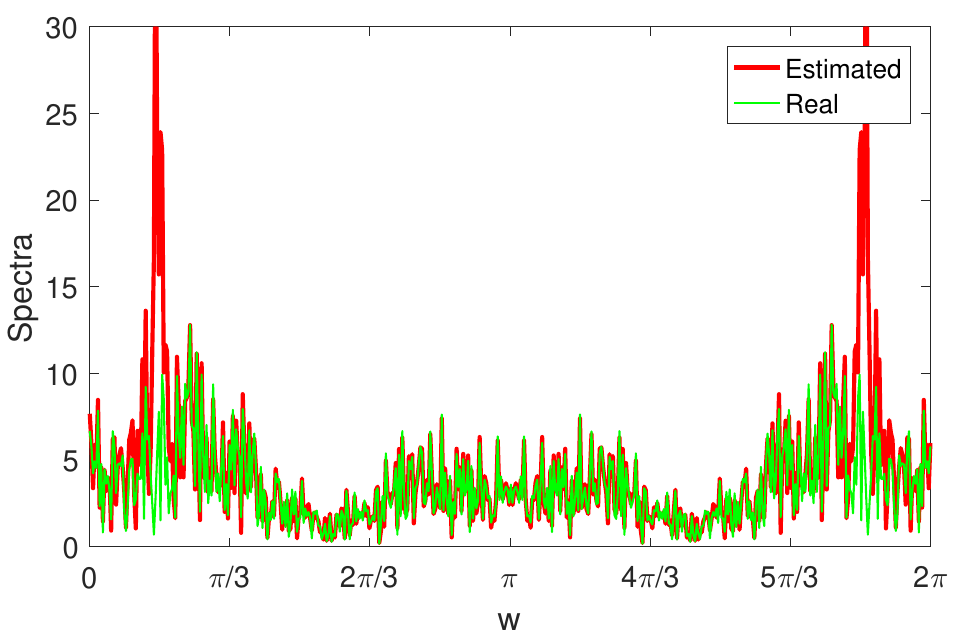}
        \end{minipage}
    }
    \subfigure[Estimated spectra of $x^0$ using $12$-lag seasonal difference]{
        \begin{minipage}[htbp]{0.45\linewidth}
            \centering
            \includegraphics[height=4cm]{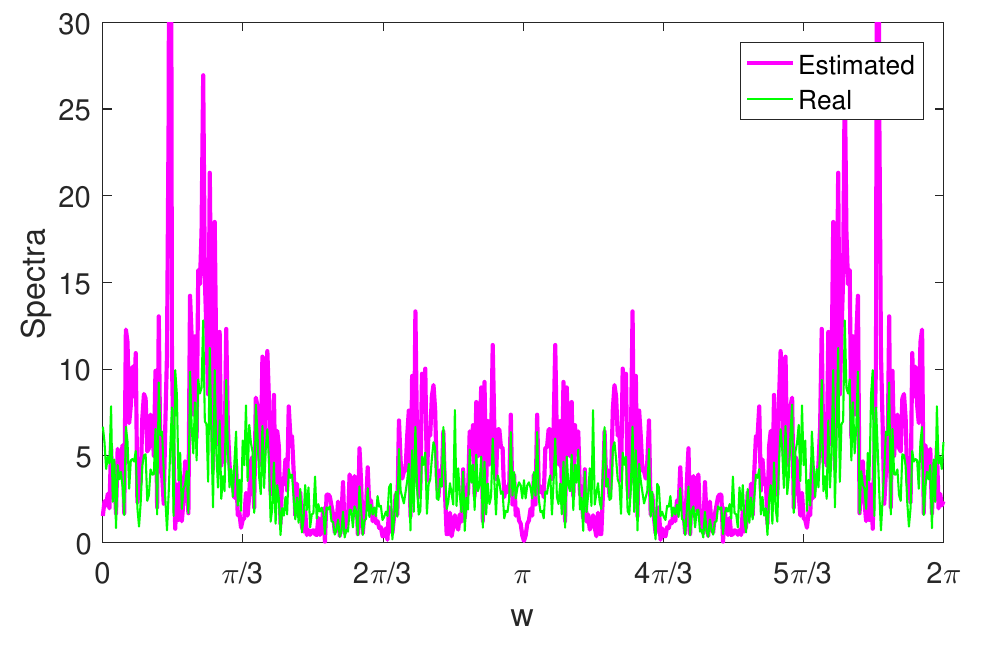}
        \end{minipage}
    }
    \centering
    \caption{Spectra analysis for the case of using SARIMA.}
    \label{fig:case3-detail}
\end{figure}

This case is concerned with the periodic time series which is suitable for SARIMA. The time series is shown in Figure \ref{fig:warmingup-case3} (c). In Figure \ref{fig:case3-detail}, we can check the spectra of the time series. We can see that there is an outstanding line-spectra (i.e., impulse, outlier) around $w = 0.5\pi$. Thus we want to design a band-stop digital filter to deny the outstanding frequency component. Since we use the Elliptic method to design a IIR filter (the function \textit{fdesign.bandstop} in MATLAB, for more, see its reference page in \cite{fdesignbandstop}), we set the parameters in Table \ref{tab:warmingup-case3-parameters}.
\begin{table}[htbp]
    \centering
    \caption{The filter parameters for the case of using SARIMA}
    \label{tab:warmingup-case3-parameters}
    \begin{tabular}{lclll}
        \toprule
                    & Value          & Comments \\
        \hline
         Fpass1     & 0.158          & First Passband Frequency ($\times \pi$)\\
         Fstop1     & 0.16          & First Stopband Frequency ($\times \pi$)\\
         Fstop2     & 0.165          & Second Stopband Frequency ($\times \pi$)\\
         Fpass2     & 0.168          & Second Passband Frequency ($\times \pi$)\\
         Apass1     & 1          & First Passband Ripple (dB)\\
         Astop      & 20          & Stopband Attenuation (dB)\\
         Apass2     & 1          & Second Passband Ripple (dB)\\
        \bottomrule
    \end{tabular}
\end{table}

This band-stop filter actually defines the system (\ref{eq:warmingup-case3}).

The spectral analysis results are given in Figure \ref{fig:case3-detail}.

The spectra of the $12$-lag seasonal difference operator could be found in Figure \ref{fig:L-lag-operator} as a comparison to Figure \ref{fig:case3-detail}.

From Figure \ref{fig:case3-detail}, we can see the $12$-lag seasonal difference operator significantly distorts the spectra at and around $w=2k\pi/12,k=0,2,3,...,11$, although it is powerful to wipe out the frequency at $w = 2\pi/12~(k = 1)$. This is why our ARMA-SIN outperforms ARIMA.

\subsection{The Case of Directly Estimating the Seasonal Component}
In this section we will use the information presented in spectra of a time series to directly estimate the seasonal (impulses in frequency domain) component. The philosophy is based on DFT (discrete Fourier transform) and Inverse DFT \cite{diniz2010digital,yang2009signals,chaparro2018signals}.

Suppose the outstanding value of FFT of $x(n)$ is at $n=K$ ($K \in \{1,2,3,...,length(\bm n)\}$) and its value is $H_{K}$, then we have $w = 2\pi K/N$. Besides, by DFT (and/or FFT), the amplitude in time domain is given as $A = 2|H_{K}|/N$ and phase is as $\varphi = \varphi({H_{k}})$, where $|H_{K}|$ denotes the modulus of $H_{K}$ and $\varphi({H_{k}})$ denotes the argument (angle) of $H_{K}$. Note that $H_{K}$ is complex-valued. Therefore the line-spectra (impulse in spectra) component of $x(n)$ with frequency of $w = 2\pi K/N$ has its expression in time domain as
\begin{equation}\label{eq:case4-detail-inv-DFT}
  x_p(n) = A\cos(wn + \varphi).
\end{equation}

\begin{figure}[htbp]
    \centering
    \includegraphics[height=4cm]{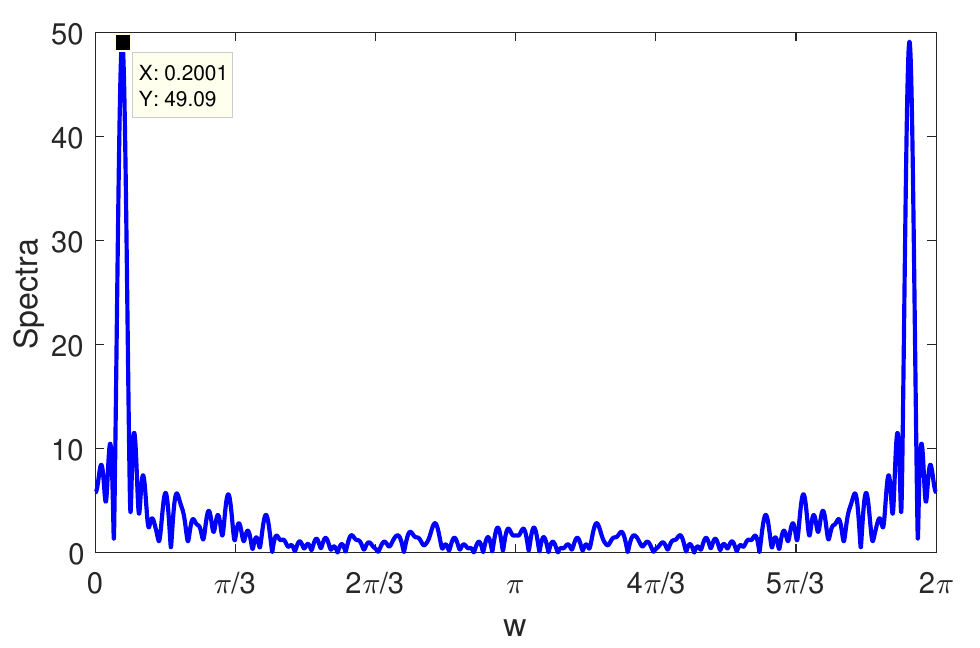}
    \centering
    \caption{Spectra analysis for the case of direct estimation.}
    \label{fig:case4-detail}
\end{figure}

Specifically in this section, the spectra of the interested time series is as Figure \ref{fig:case4-detail}.

From the figure we can directly know that $w = 0.2001$, and $A = 2 \times 49.09/N = 0.9721$ ($N=101$). Besides, by checking the corresponding phase at this frequency $w = 0.2001$, we have $\varphi = -1.5440$. Thus (\ref{eq:case4-detail-inv-DFT}) admits (\ref{eq:warmingup-case4}).

\section{Conclusions}
We have in this paper discussed the natures, philosophies, effectiveness, insufficiencies and improvements of ARIMA and SARIMA from the perspectives of Linear System Analysis, Spectra Analysis and Digital Filtering. We show that S-ARIMA also distorts desired frequencies when making stationary a time series. Fortunately, our ARMA-SIN could remain the innocent frequencies unchanged. However, we should admit that although the displayed ARMA-SIN is powerful, general and interesting, it is relatively hard for beginners to make use of, compared to the simple ARIMA, SARIMA and the like. Applying this method requires enough experience to identify the  spectral components of interest and setup proper parameters like cut-off frequencies to extract or discard them by designing a proper digital filter. Although depressive mentioning this, we should make clear that it is still a bright future if we can design some powerful algorithms to help us automatically select the proper parameters to design the useful digital filters, just like the cross-validation method in training a LASSO model. However, this challenging work should be jointly handled with inspired scholars in this area.

\appendix
In order not to confuse readers from different communities, we mention some basic glossaries of interest in this paper.
\begin{itemize}
\item ARMA: Autoregressive Moving Average;
\item ARIMA: Autoregressive Integrated Moving Average;
\item SARIMA: Seasonal Autoregressive Integrated Moving Average;
\item Spectrum: The spectrum of a time function is its Fourier transform which describes a time series in Fourier frequency domain;
\item Spectral Analysis: The analysis and processing for a time series in Fourier frequency domain. The Fourier transform and inverse Fourier transform connect the time domain and Fourier frequency domain;
\item (Discrete) Linear System: A discrete linear operator, defined by a difference equation (see (\ref{eq:ARMAPara})) in autoregressive moving average form, that transforms a time series into another time series (for example, the difference operator);
\item Digital Filter: A linear system that takes effect of transforming a time series in Fourier frequency domain. However, it works for a time series in time domain. For example, the moving average method, although it is directly applied in time domain, in essence, it removes the low-frequency component (i.e., the trend) in Fourier frequency domain;
\item SADFA: Spectral Analysis and Digital Filtering Approach used to analyze and process a time series;
\item DFT/DFS: Discrete Fourier Transform/Series;
\item FFT: Fast Fourier Transform;
\item DTFT: Discrete Time Fourier Transform;
\end{itemize}
Besides, two terms Time Series and Stochastic Process are used interchangeably in this paper.

\bibliographystyle{IEEEtran}
\bibliography{References.bib} 

\end{document}